\documentclass[pra,twocolumn,showpacs,floatfix,showkeys]{revtex4-1}
\usepackage{times,amsmath,amssymb,amstext,latexsym,float,graphicx,color}
\usepackage[dvipsnames]{xcolor}

\usepackage[hidelinks]{hyperref}
\hypersetup{colorlinks=true, citecolor=blue, urlcolor=blue, linkcolor=blue}

\begin{document}

\title{Breathing mode in two-dimensional binary self-bound Bose gas droplets}
\author{P. St\"urmer}
\email[]{philipp.sturmer@matfys.lth.se}
\affiliation{Mathematical Physics and NanoLund, Lund University, Box 118, 22100 Lund, Sweden}
\author{M. Nilsson Tengstrand}
\affiliation{Mathematical Physics and NanoLund, Lund University, Box 118, 22100 Lund, Sweden}
\author{R. Sachdeva}
\affiliation{Mathematical Physics and NanoLund, Lund University, Box 118, 22100 Lund, Sweden}
\author{S. M. Reimann}
\affiliation{Mathematical Physics and NanoLund, Lund University, Box 118, 22100 Lund, Sweden}

\date{\today}

\begin{abstract}
In this work, we present the study of the stationary structures and the breathing mode behavior of a two-dimensional self-bound binary Bose droplet. We employ an analytical approach using a variational ansatz with a super-Gaussian trial order parameter and compare it with the numerical solutions of the extended Gross-Pitaevskii equation. We find that the super-Gaussian is superior to the often used Gaussian ansatz in describing the stationary and dynamical properties of the system. We find that for sufficiently large non-rotating droplets the breathing mode is energetically favourable compared to the self-evaporating process. For small self-bound systems our results differ based on the ansatz. Inducing angular momentum by imprinting multiply quantized vortices at the droplet center,  this preference for the breathing mode  persists independent of the norm.
\end{abstract}

\maketitle

\section{Introduction\label{ch:Introduction}}
For bound systems to form without external confinement, a balance between repulsive and attractive forces is required, this principle holds for systems ranging from water droplets to atomic nuclei and metallic clusters. 
For ultra-cold bosonic atoms the possibility to form self-bound droplets was proposed for binary gases with suitably tuned contact interactions~\cite{ref:contact-drop_bulgac,ref:contact-drop_3dim_Petrov}, where higher-order terms in the total energy density functional may become sizeable. 
In fact, such Bose droplets were first discovered for dipolar bosonic systems~\cite{ref:dipole-drop_Kadau-Pfau,ref:dipole-drop_schmitt,ref:dipole-drop_Barbut-Pfau,ref:dipole-drop_chomaz,ref:dipole-drop_Wenzel,ref:dipole-drop_scissors-mode_ferrier-barbut}, 
where the effective interatomic  interactions may be adjusted such that self-bound states become stabilized by quantum fluctuations~\cite{ref:dipole-drop_Saito,ref:dipole-drop_wachtler1,ref:dipole-drop_baillie1,
ref:dipole-drop_bisset1,ref:dipole-drop_wachtler2,ref:dipole-drop_chomaz,ref:dipole-drop_baillie2}. 
The experimental discovery of droplets in dipolar gases was soon followed by their realization in binary Bose 
gases~\cite{ref:contact-drop_Semeghini,ref:contact-drop_Cabrera}, implementing Petrov's original proposal~\cite{ref:contact-drop_3dim_Petrov}. 
The quantum fluctuation contributions to the total energy density functional are effectively represented by the 
Lee-Huang-Yang (LHY) terms~\cite{ref:LHY-1957},  extending the  usual mean field (MF)  Gross-Pitaevskii equation (see \cite{ref:book_pethick-smith,ref:book_stringari-pitaevskii} for a review). Usually, the LHY terms can be neglected as their contribution to the total interaction energy is negligible. However, in the case of a binary condensate where the MF interactions between components are tuned such that they become very small, the LHY contributions may dominate the system's properties~\cite{ref:contact-drop_3dim_Petrov}. In a binary three-dimensional condensate the LHY terms are positive definite~\cite{ref:LHY-1957} and thus purely repulsive. A three-dimensional binary condensate on the verge of collapse in the MF description can thus be stabilized via the repulsive LHY terms. In lower dimensions, however, the corrections can take on an attractive or repulsive form~\cite{ref:contact-drop_low-dim_Petrov-Astrakharchik, ref:LHY-q2d_Zin,ref:LHY-q2d_Ilg}.\\
\noindent The proposal and realization of self-bound Bose gas droplets sparked recent research efforts to set focus on different aspects such as vorticity embedded in dipolar \cite{ref:dipole-drop-vortices_Cidrim,ref:dipole-drop-vortex_Lee} or contact-interacting systems \cite{ref:contact-drop_vortex_li,ref:contact-drop_vortex_mikael,ref:contact-drop_kartashov1,ref:contact-drop_kartashov2,ref:contact-drop_rot_kavoulakis}, supersolid behavior in dipolar and spin-orbit coupled systems~ \cite{ref:supersolid_ancilotto,ref:supersolid_boettcher,ref:supersolid_chomaz,ref:supersolid_mikael-david,ref:supersolid_rashi,ref:supersolid_roccuzzo1,ref:supersolid_roccuzzo2,ref:supersolid_tanzi_1,ref:supersolid_tanzi2} or collective excitations \cite{ref:col-ex-drop_astrakharchik,ref:contact-drop-sol-cross_cappelaro,ref:contact-drop_one-dim_superGauss,ref:contact-drop_col-ex_BdG_astrakharchik, ref:col-ex_3D_huhui}.
For trapped gases, the elementary excitation modes have been of fundamental interest ever since the celebrated experimental realizations of atomic Bose-Einstein condensates~\cite{ref:exp-bec_Anderson,ref:exp-bec_Bradley,ref:exp-bec_Davis}: Examples range from a hydrodynamic description \cite{ref:trap_col-ex_Stringari} to the Bogoliubov-de Gennes equations \cite{ref:bog-deGen_edwards, ref:book_pethick-smith, ref:book_stringari-pitaevskii} and to solving the Euler-Lagrange equations \cite{ref:trap_variat_perez2,ref:trap_variat_perez1,ref:var_salasnich,ref:trap_var_bec-td_castin}.   
Breathing mode oscillations were discussed in terms of the underlying symmetry properties of a Bose gas in two dimensions~\cite{ref:breath_pitaevskii}, and semiclassical methods have been probed to describe low-lying excitations in the limit of large boson numbers~\cite{ref:col-ex_Sinha}.  Collective excitations of BECs carrying vortices or vortex lattices have also been studied extensively; to mention only a few examples from the vast literature, see  
Refs~\cite{ref:breath_pitaevskii,ref:col-ex_Sinha,ref:var_normal-modes-vortex_fetter1,ref:var_normal-modes-vortex_fetter2,ref:col-ex-vortex-lattice_mizushima,ref:breath-LLL_watanabe,ref:kelvin-waves_Simula,ref:col-ex-vortices_simula,ref:trap_variat-release_Teles,ref:col-ex-vortex_teles}. 
More recently and in the context of Bose droplets, in Ref. \cite{ref:LHY_fluid_bruun}  a variational approach for 
solving the Euler-Lagrange equations was applied to calculate the breathing mode in a harmonically confined three-dimensional Bose gas where the MF interactions are completely cancelled, such that the LHY interaction is effectively the only interaction acting on the gas, creating a so-called LHY fluid.\\
Excitations compete with the self-evaporative process, naturally occurring in self-bound SBose gases due to a non-positive chemical potential $\mu\leqslant 0$, and are generally only observable if the associated excitation energy is lower than $|\mu|$. However, this requirement can stay unfulfilled for self-bound systems~\cite{ref:contact-drop_3dim_Petrov} depending on the system's dimensionality, if the excitation energy exceeds $|\mu|$, see 
Refs.~\cite{ref:contact-drop_col-ex_BdG_astrakharchik,ref:contact-drop_col-ex_BdG_astrakharchik,ref:contact-drop-sol-cross_cappelaro}.\\
Recent works on collective excitations in self-bound systems have mainly focused on one-dimensional droplets~\cite{ref:col-ex-drop_astrakharchik,ref:contact-drop_low-dim_Petrov-Astrakharchik} or on the cross-over between the soliton and droplet regime~\cite{ref:col-ex-drop_astrakharchik}. Besides solving the Bogoliubov-de Gennes equations in 
one~\cite{ref:contact-drop_col-ex_BdG_astrakharchik} and in three dimensions~\cite{ref:contact-drop_3dim_Petrov, ref:col-ex_3D_huhui}, a Gaussian trial order parameter is often utilized to describe the properties of the droplet state~\cite{ref:col-ex-drop_astrakharchik,ref:contact-drop-sol-cross_cappelaro, ref:col-ex_3D_huhui}. However, due to the self-bound nature and the resulting flat-top profile of the droplet, the classical Gaussian shape proves to be a poor approximation of stationary properties of droplets. Yet the Gaussian shape still represents the dynamics reasonably well in one dimension \cite{ref:col-ex-drop_astrakharchik}, although overestimating the results in three dimensions \cite{ref:col-ex_3D_huhui}. In search for a better approximation of the order parameter, a $\sqrt{1-\tanh(x)}$ function in the Thomas-Fermi limit has been utilized~\cite{ref:contact-drop_rot_kavoulakis}, which imposes difficulties in an analytical variational ansatz. To overcome the limitations by these functions, we here study the breathing mode and stationary properties using a super-Gaussian ansatz for droplets in two dimensions, following  Refs.~\cite{ref:contact-drop_one-dim_superGauss, ref:sg-solitons_baizakov} for one-dimensional systems. The super-Gaussian allows for a more precise description of a flat-top shape than the pure Gaussian, while still allowing for an easier analytical treatment compared to a $\sqrt{1-\tanh(x)}$ function.\\

\noindent In continuation of our previous work~\cite{ref:contact-drop_vortex_mikael}, in this paper we investigate the lowest-lying collective excitations of self-bound Bose-gas droplets in two dimensions. 
These collective excitations are generally referred to as (monopole) breathing modes, where the radial extent of the droplet oscillates unidirectionally. The work is structured as follows. In Sec.~\ref{ch:model} we introduce the basic model for a two-dimensional binary self-bound Bose gas droplet. In  
Sec.~\ref{ch:variational_analysis} we use a super-Gaussian function as a trial order parameter and solve the Euler-Lagrange equations for complex fields of droplets. The calculation is followed by an analysis and comparison of stationary properties to numerical results self-bound systems in Sec.~\ref{ch:stationary_properties} and the breathing frequency in 
Sec.~\ref{ch:dynamic_properties}. We continue in Sec.~\ref{ch:numerical_results} by studying droplets carrying angular momentum in the form of a phase-imprinted singly- or multiply-quantized vortex at the droplet center.  We analyze the breathing mode frequency and energetically compare the mode with the self-evaporation process. 


\section{Model\label{ch:model}}
\noindent Let us now consider a binary Bose-Einstein condensate in two dimensions,  and  label the two components by the indices $1$ and $2$.  The atoms are chosen to have equal mass and equal intra-component scattering lengths $a_{11} = a_{22} = a$, with an inter-component scattering length denoted by $a_{12}$ . The number of atoms in each component is set to be equal and consequently one also has equal atom densities $n_{1} = n_{2} = n$.  
The two-component system thus reduces to effectively only one component,  
where the MF interaction cancels and only the LHY terms beyond mean field remain, and the   
time-dependent extended Gross-Pitaevskii equation (eGPe) takes the form 
\begin{align}
    \begin{split}
        i\frac{\partial\psi}{\partial t} =& \Big(-\frac{1}{2}\nabla^2 +|\psi|^2\ln|\psi|^2\Big)\psi ~, \label{eq:re:td-GPe}
    \end{split}
\end{align}
see Ref.~\cite{ref:contact-drop_low-dim_Petrov-Astrakharchik}. 
Here, $\psi =\psi(\mathbf{r},t)$ is the order parameter with $\mathbf{r} = (x,y)$. As in Refs.~\cite{ref:contact-drop_low-dim_Petrov-Astrakharchik,ref:contact-drop_vortex_mikael,ref:contact-drop_vortex_li,ref:contact-drop_kartashov2,ref:supersolid_rashi}, the scaling invariances have been used to bring the system into dimensionless form, such that it conveniently only depends on the norm $N = \int d^2\mathbf{r}|\psi(\mathbf{r})|^2$. Given Eq.\,(\ref{eq:re:td-GPe}) the energy density $\mathcal{E}[\psi]$ of the self-bound system is then given by
\begin{align}
    \mathcal{E}[\psi] = \frac{1}{2}|\nabla\psi|^2 + \frac{1}{2}|\psi|^4\ln{\frac{|\psi|^2}{\sqrt{e}}}. \label{eq:re:energydensity}
\end{align}{}\noindent 
For the comparision of the variational ansatz discussed in the following, Eq.\,(\ref{eq:re:td-GPe}) is solved numerically using the Fourier split-step method in imaginary and real time.

\section{Super-Gaussian Variational Ansatz}\label{ch:variational_analysis}

\begin{figure}
    \centering
    \includegraphics[width = 0.45\textwidth]{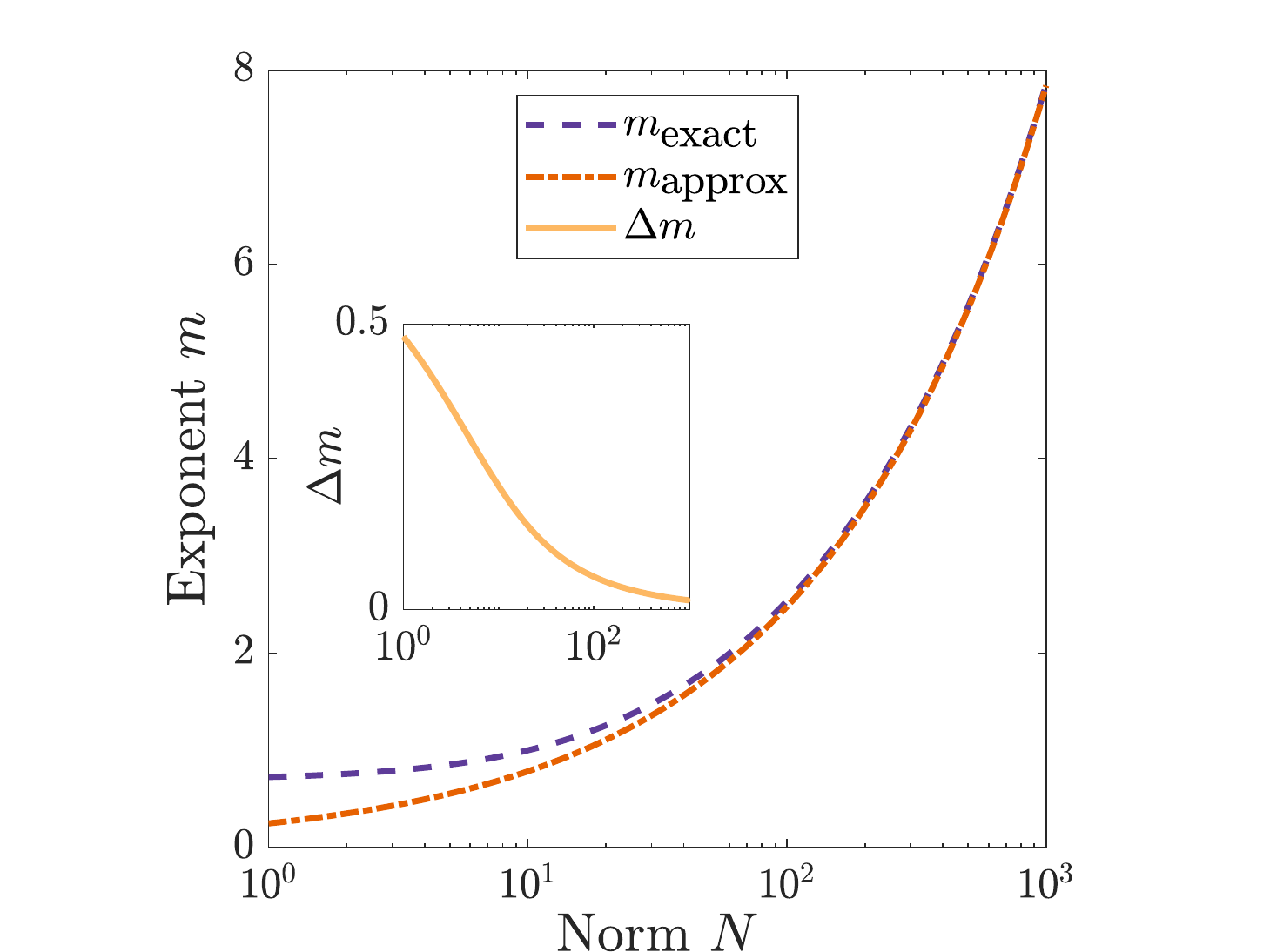}
    \caption{Optimal value for the exponent $m$ of the super-Gaussian ansatz with increasing norm $N$. Solved numerically for Eq.\,(\ref{eq:var:m_exact}) \emph{(dashed line)} and approximated for large $N$ according to Eq.\,(\ref{eq:var:m_approx}) \emph{(dotted-dashed line)}. The inset shows the difference $\Delta m = m_{\text{exact}}-m_{\text{approx}}$. With increasing $N$ the error $\Delta m$ approaches zero, suggesting that the analytical approximation is suitable for flat-top droplets.}
    \label{fig:no_rot_m}
\end{figure}

In this section we use a super-Gaussian trial order parameter to calculate the stationary and dynamic properties of a self-bound binary droplet by taking advantage of the Euler-Lagrange equations for a complex field. 
We then continue by comparing the analytical solution to the 
numerical solution of the eGPe in Eq.\,(\ref{eq:re:td-GPe}).
Recent works which made use of a variational analysis of such self-bound states used either a Gaussian ansatz  for one or three-dimensional droplets~\cite{ref:col-ex-drop_astrakharchik,ref:contact-drop-sol-cross_cappelaro, ref:col-ex_3D_huhui}, or a $\sqrt{1-\tanh{x}}$ ansatz in the Thomas-Fermi approximation for two-dimensional droplets~\cite{ref:contact-drop_rot_kavoulakis}. While the Gaussian ansatz for the breathing frequency yields small errors compared to the numerical eGPe solution for one-dimensional problems~\cite{ref:col-ex-drop_astrakharchik}, it has limitations to properly describe the spatial properties of self-bound systems as it lacks the characteristic flat-top shape for high norms. Furthermore it overestimates the breathing mode frequency in three dimensions~\cite{ref:col-ex_3D_huhui}. On the other hand, a $\sqrt{1-\tanh{x}}$ function accurately describes the spatial properties, but faces limitations to describe the system for small norms~\cite{ref:contact-drop_rot_kavoulakis} and also becomes analytically cumbersome when treated with the Euler-Lagrange equations. In comparison, a super-Gaussian ansatz offers a good middle ground, as it can be analytically handled and describes the low- and high-norm systems properly \cite{ref:sg-solitons_baizakov,ref:contact-drop_one-dim_superGauss,ref:sg-opt_Karlsson,ref:sg-opt_Tsoy}. For the following calculation we thus use the super-Gaussian as a trial order parameter. Assuming a circular droplet shape, we write   
\begin{align}
    \psi (r,t)= A(t)\exp\left(ib(t)r^2-\frac{1}{2}\left(\frac{r}{R(t)}\right)^{2m}\right)~\label{eq:var:supergaussian}
\end{align}{}\noindent 
with radial coordinate $r$, real amplitude $A(t)$,  chirp $b(t)$ and width $R(t)$.  The real and positive exponent $m$  is determined as a function of the norm $N$. \\
The dynamics of a complex field, such as the reduced two-dimensional two-component Bose gas for some generalized coordinate $q_i=\{A(t),b(t),R(t)\}$, emerges from the respective Euler-Lagrange equations
\begin{align}
    \frac{\partial \mathfrak{L}}{\partial q_i} = \frac{d}{dt}\frac{\partial \mathfrak{L}}{\partial\dot{q_i}},\label{eq:var:Euler-Langrange}
\end{align}{}\noindent
where the Lagrangian $\mathfrak{L}$ is calculated by averaging the Lagrangian density $\mathcal{L}$ over the full space
\begin{align}
    \mathfrak{L} =\langle\mathcal{L}\rangle = \int d^2\mathbf{r}\,\mathcal{L}\,.\label{eq:var:Lagrangian}
\end{align}
The Lagrangian density $\mathcal{L}$ for the two-dimensional droplet system results from the energy density for a self-bound system in Eq.\,(\ref{eq:re:energydensity}) as
\begin{align}
    \begin{split}
        \mathcal{L}[\psi,\psi^*] = \frac{i}{2}\left(\psi^*\frac{\partial\psi}{\partial t}-\psi\frac{\partial\psi^*}{\partial t}\right)-\mathcal{E}[\psi].\label{eq:var:Lagrangiandensity}
    \end{split}
\end{align}{}\noindent 
Normalizing the wave function gives the number of particles in the condensate,
\begin{align}
    N = \int d^2\mathbf{r}\,|\psi |^2 = \frac{\pi}{m}\Gamma\left(\frac{1}{m}\right)A^2R^2\label{eq:var:norm_SG},
\end{align}{}\noindent 
where $\Gamma(m)$ is the gamma function. We continue to write $\Gamma\left(\frac{1}{m}\right) = \Gamma_1$, $\Gamma\left(\frac{2}{m}\right) = \Gamma_2$ and $\Gamma\left(\frac{2}{m}\right)/\Gamma\left(\frac{1}{m}\right) = \Gamma$. After evaluation of the integral in Eq.\,(\ref{eq:var:Lagrangian}) by inserting the trial wavefunction in Eq.\,(\ref{eq:var:Lagrangiandensity}) and eliminating powers of $A$ in favor of the norm $N$, the Lagrangian reads
\begin{align}
    \begin{split}
        \frac{\mathfrak{L}}{N}=&-\dot{b}R^2\Gamma-2b^2R^2\Gamma-\frac{m^2}{2R^2\Gamma_1}-\\
        &\frac{Nm}{2\pi R^2}\frac{2^{-1/m}}{\Gamma_1}\ln\left(\frac{Nm}{\pi R^2}\frac{1}{\Gamma_1}\right)+\frac{N}{4\pi R^2}\frac{2^{-1/m}}{\Gamma_1}\left(m+1\right).\label{eq:var:Lagrangian_SG}
    \end{split}
\end{align}{}\noindent 
Thus, the Euler-Lagrange equations for the variational parameters $b(t)$ and $R(t)$ become
\begin{align}
    b &= \frac{1}{2}\frac{\dot{R}}{R},
\end{align}
and
\begin{align}
    \begin{split}
        \Ddot{R} =& \frac{1}{\Gamma_2}\Bigg[\frac{m^2}{R^3}+\frac{N}{2\pi R^3} 2^{-1/m}(m-1)+\\
        &\frac{Nm}{\pi R^3}2^{-1/m}\ln\left(\frac{Nm}{\pi R^2}\frac{1}{\Gamma_1}\right)\Bigg]= -\frac{dU_{\textrm{eff}}}{dR}.\label{eq:var:EL_SG}
        \end{split}
\end{align}{}\noindent
Let us now first determine $m$ as a function of $N$. We extremize the Lagrangian $\mathcal{L}$ with respect to $m$, which gives the transcendental equation with the analytical exact value $m_{\textrm{exact}}$
\begin{align}
    0 = \frac{m_{\textrm{exact}}}{2}(\ln{2}-m_{\textrm{exact}})+\frac{N}{4\pi}2^{-1/m_{\textrm{exact}}}(2\ln{2}-1).\label{eq:var:m_exact}{}
\end{align}{}
For large $N$ we have an approximation $m_{\textrm{approx}}$
\begin{align}
    m_{\textrm{approx}}\approx\sqrt{\frac{N}{2\pi}(\ln{4}-1)}.\label{eq:var:m_approx}
\end{align}{}\noindent
In Fig. \ref{fig:no_rot_m} we plot $m$ according to Eqs.\,(\ref{eq:var:m_exact}) and (\ref{eq:var:m_approx}) as dashed and dotted-dashed lines respectively, as well as the difference $\Delta m = m_{\textrm{exact}} - m_{\textrm{approx}}$ (see inset). 
The difference $\Delta m$ approaches $0$ as $N\rightarrow\infty$, so 
that we can use  the approximate expression in  Eq.\,(\ref{eq:var:m_approx}) for droplets with large $N$.
Looking at the minimum of the effective potential $U_{\text{eff}}$, as defined in Eq.\,(\ref{eq:var:EL_SG}), gives us an expression for the maximum density $n_m$ and the radial extent of the condensate $R_0$, such that
\begin{align}
    A^2|_{t=0} = n_m = \exp\left[\frac{1}{2m}(1-m)-m\pi\frac{2^{1/m}}{N}\right].\label{eq:var:equilbrium-density_SG}
\end{align}{}\noindent
For large norms the maximum density approaches $1/\sqrt{e}$ as expected for this choice of scaling \cite{ref:contact-drop_vortex_mikael,ref:contact-drop_vortex_li,ref:contact-drop_kartashov1,ref:supersolid_rashi}. The mean radius $\sqrt{\langle r^2\rangle}$ depends on the width of the cloud,
\begin{align}
    \langle r^2\rangle = \Gamma R^2,\label{eq:var:radius_expectation_value_SG}
\end{align}{}\noindent
which simply reduces to the Gaussian case $\langle r^2\rangle = R^2$ for $m=1$. 
Furthermore, the width of a stationary condensate is 
\begin{align}
        R_0^2   = \frac{N}{\pi n_m}\frac{m}{\Gamma_1}.\label{eq:var:radial_extent_SG}
\end{align}{}\noindent
The low-lying excitation frequencies are found by assuming a single sinusoidal oscillation in $R(t)$ with frequency $\omega_0$, such that the breathing frequency $\omega_0$ is given by
\begin{align}
    \frac{d^2U_\textrm{eff}}{dR^2} = \omega_0^2\, , \label{eq:var:freq_eigenval}
\end{align}{}\noindent
and one obtains
\begin{align}
    \omega^2_0 = n_m^2\frac{\Gamma_1}{\Gamma}\frac{2\pi}{Nm}2^{-1/m}\, . \label{eq:var:freq_SG}
\end{align}{}\noindent
The result in Eq.~(\ref{eq:var:freq_SG}) can be divided into a low and high norm $N$ regime. In the low norm regime, the behavior is dominated by the exponential behavior of $n_m$, which approaches a constant value for high norm, such that the $1/N$ dependency governs the breathing frequency's behavior.

From the time-independent eGPe, Eq.\,(\ref{eq:re:td-GPe}), we obtain the chemical potential $\mu$ as
\begin{align}
    \mu = -\frac{n_m}{2}\left(\frac{m\pi}{N}+2^{-1/m}\right)\label{eq:var:chemical-pot_SG}
\end{align}\noindent
With $N\rightarrow\infty$, $\mu$ then approaches $-1/(2\sqrt{e})$ as suggested by Refs. \cite{ref:contact-drop_low-dim_Petrov-Astrakharchik,ref:contact-drop_vortex_li}.


\subsection{Stationary properties}\label{ch:stationary_properties}

\begin{figure}
    \centering
    \includegraphics[width = 0.45\textwidth]{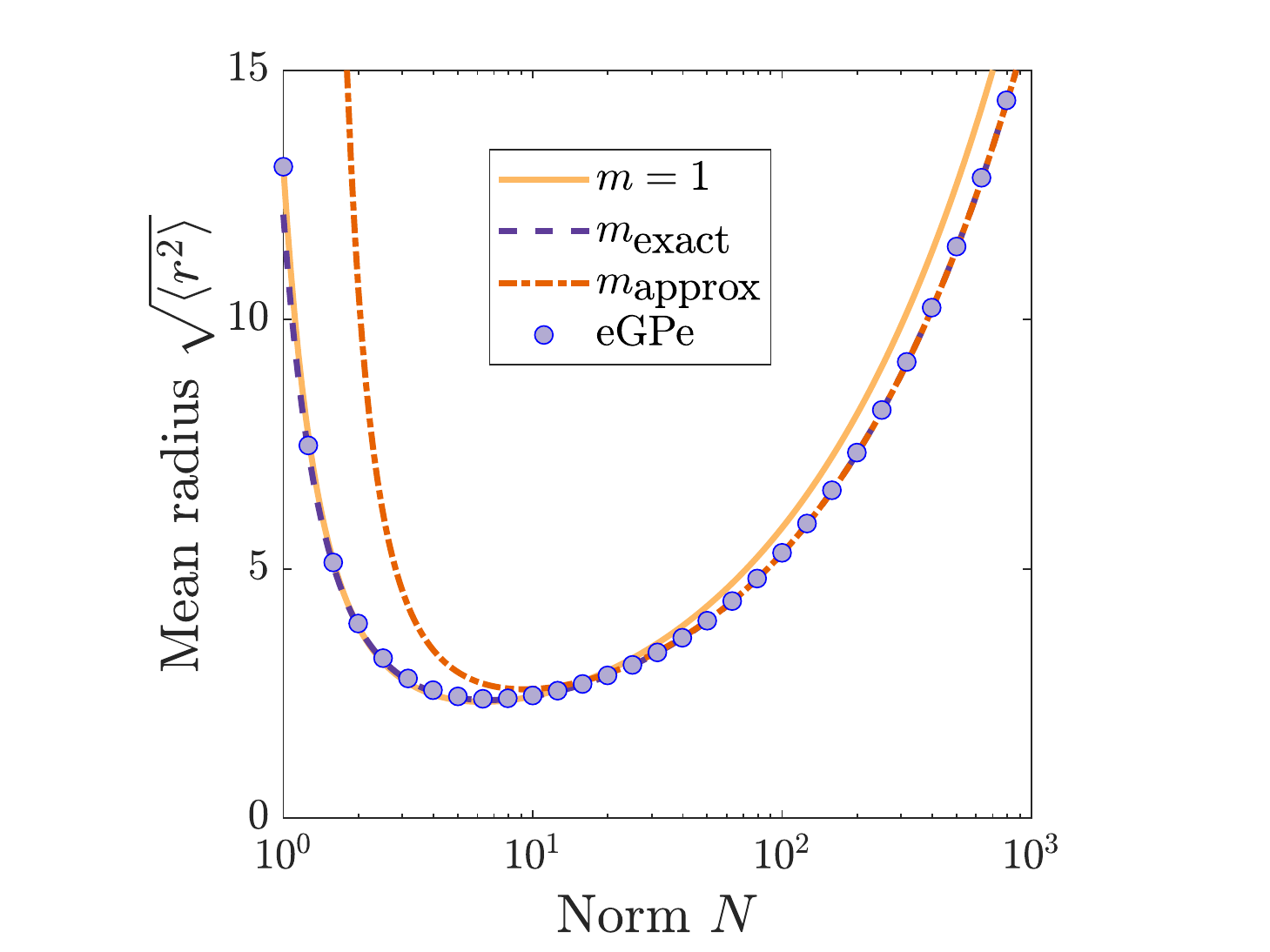}
    \caption{Mean radius $\sqrt{\langle r^2\rangle}$ of the self-bound droplet based on different approaches, as indicated by the legend. While the Gaussian trial order parameter ($m=1$) is a good approximation for small $N$, it fails to describe the mean radius accurately for  larger norm.}
    \label{fig:no_rot_r_square}
\end{figure}

\noindent Let us now use the previously obtained expressions for the exact and approximative value of $m$ as well as the Gaussian case $(m=1)$ and compare them to the numerical solution of the eGPe equation, Eq.\,(\ref{eq:re:td-GPe}). 
In the following, unless otherwise specified, we will use solid lines for the Gaussian case with $m=1$, dashed lines and dotted-dashed lines for $m$ according to Eq.\,(\ref{eq:var:m_exact}) and (\ref{eq:var:m_approx}), respectively, and numerical results of the eGPe in bullets for norms $1\le N\le 1000$.\\
We start by comparing the mean radius $\sqrt{\langle r^2\rangle}$ in Fig. \ref{fig:no_rot_r_square}. As expected from our previous discussion about $m$, the exact and approximate descriptions perfectly overlap for large $N$. With decreasing $N$ the approximate description starts to deviate and our variational approach is well approximated by the Gaussian form of the order parameter for $m=1$. For all values of $N$ the numerical results obtained by solving the eGPe coincide directly with $m_{\textrm{exact}}$ solution. For the case of a large norm, it follows the behavior expected of a growing droplet. For a smaller norm, however, the system's properties are dominated by the kinetic energy.
This bears some similarity to the bright soliton-like structures known to occur in one-dimensional systems \cite{ref:sol_drop_cheiney,ref:sol_carr,ref:sol_garcia,ref:sol_salasnich}.\\

\begin{figure}
    \centering
    \includegraphics[width = 0.45\textwidth]{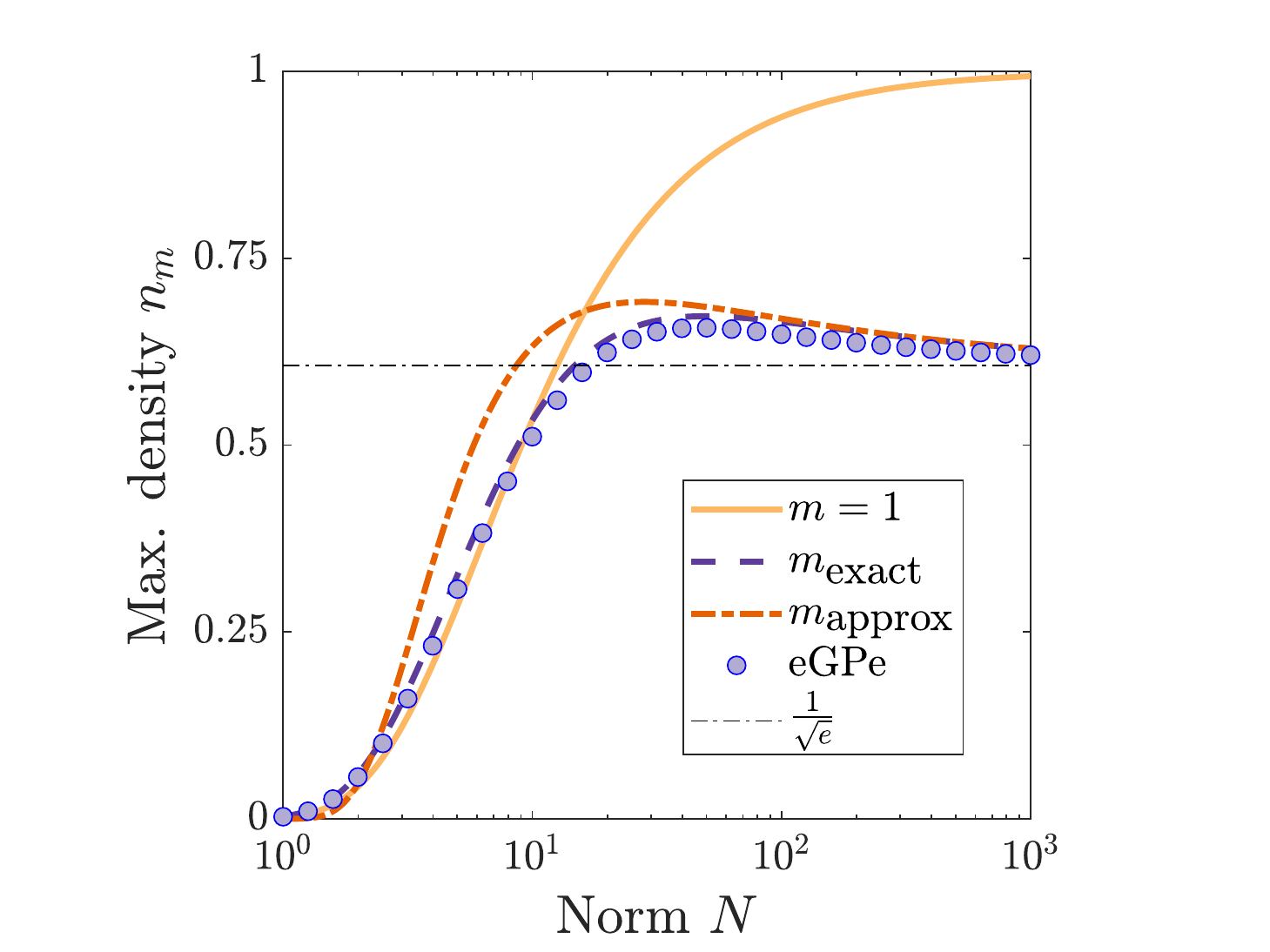}
    \caption{Maximum density $n_m$ of the droplet based on the variational and numerical solutions, as indicated in the legend. As proposed in Refs. \cite{ref:contact-drop_low-dim_Petrov-Astrakharchik,ref:contact-drop_vortex_li} $n_m$ approaches $1/\sqrt{e}$ with increasing $N$ for the super-Gaussian trial order parameter in its exact and approximate form.  With increasing norm $N$ the Gaussian ($m=1$) deviates significantly. The numerical solutions of the eGPe, Eq.\,(\ref{eq:re:td-GPe}), are in good agreement with the variational results for $m_{\mathrm {exact}}$ and $m_{\mathrm{approx}}$.}
    \label{fig:no_rot_n0.eps}
\end{figure}

\begin{figure}
    \centering
    \includegraphics[width = 0.45\textwidth]{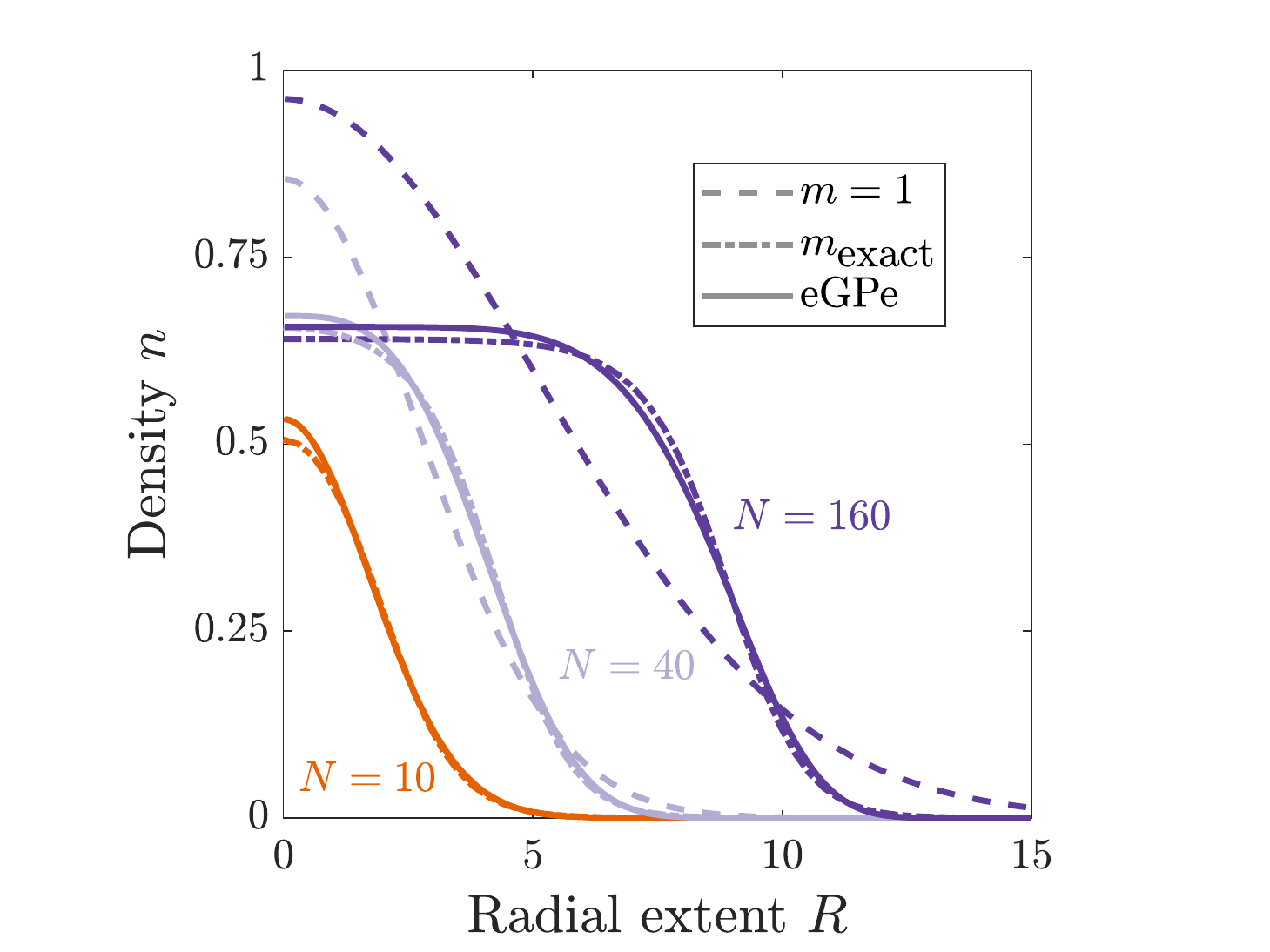}
    \caption{Density distributions $n(R)$ based on the Gaussian trial order parameter (dashed), the super-Gaussian trial order parameter with $m_{\text{exact}}$ (dotted-dashed) and a numerical solution of the eGPe (full lines) for different $N$. With increasing $N$ the super-Gaussian density distribution approaches the flat-top limit.}
    \label{fig:no_rot_density_distribution}
\end{figure}

\noindent As mentioned above, the maximum density $n_m$ approaches $1/\sqrt{e}$ and $m_{\textrm{approx}}$ in Eq.\,(\ref{eq:var:m_approx}) gives the correct behavior in the limit of large $N$, as it can be seen in 
Fig.~\ref{fig:no_rot_n0.eps}. With increasing $N$ the Gaussian with $m=1$ starts to deviate strongly from the super-Gaussian and numerical solution of the self-bound system. For small $N$, however, $m=1$ approximates the super-Gaussian reasonably well, with only small deviations, while $m_{\textrm{approx}}$ differs significantly as expected. All our numerical values lie within a $2\%$ deviation directly below the super-Gaussian with $m_{\textrm{exact}}$.\\
\noindent Having analyzed the mean radius $\sqrt{\langle r^2\rangle}$ and the maximum density $n_m$ we now look at the density profile,  comparing the exact super-Gaussian, the Gaussian and the numerical solutions of the eGPe in Fig. \ref{fig:no_rot_density_distribution}. The dashed lines represent the Gaussian with $m=1$, and the full lines the super-Gaussian. As already anticipated from the previous discussion, for small $N$ both solutions coincide with each other. For increasing $N$ the deviations increase and the super-Gaussian solution slowly approximates the characteristic flat-top shape.\\
\noindent Finally, let us look at the chemical potential $\mu$, as specified it in Eq.\,(\ref{eq:var:chemical-pot_SG}) above. Due to $\mu \leqslant 0$ in self-bound condensates, an underlying self-evaporation process exists. Thus, in principle no other mode with energy $E$ should be observable as long as $E/|\mu|\geqslant1$. In a self-bound droplet, this may not be always the case, depending on the system's dimensionality, as shown in \cite{ref:contact-drop_3dim_Petrov,ref:contact-drop_low-dim_Petrov-Astrakharchik,ref:contact-drop-sol-cross_cappelaro}. As before we compare the exact and approximate solutions for $m$, the Gaussian solution with $m=1$ and the numerical results in Fig. \ref{fig:no_rot_omega-mu}. While the exact and approximate solutions  converge towards the theoretical value of $1/(2\sqrt{e})$ for high $N$, the Gaussian solution deviates significantly. Again, the numerical solution of the eGPe,  
Eq.\,(\ref{eq:re:td-GPe}) agrees well with the variational approach both in the limits of small and large norms $N$.

\subsection{The breathing mode in a two-dimensional droplet}\label{ch:dynamic_properties}

\begin{figure}
    \centering
    \includegraphics[width=0.45\textwidth]{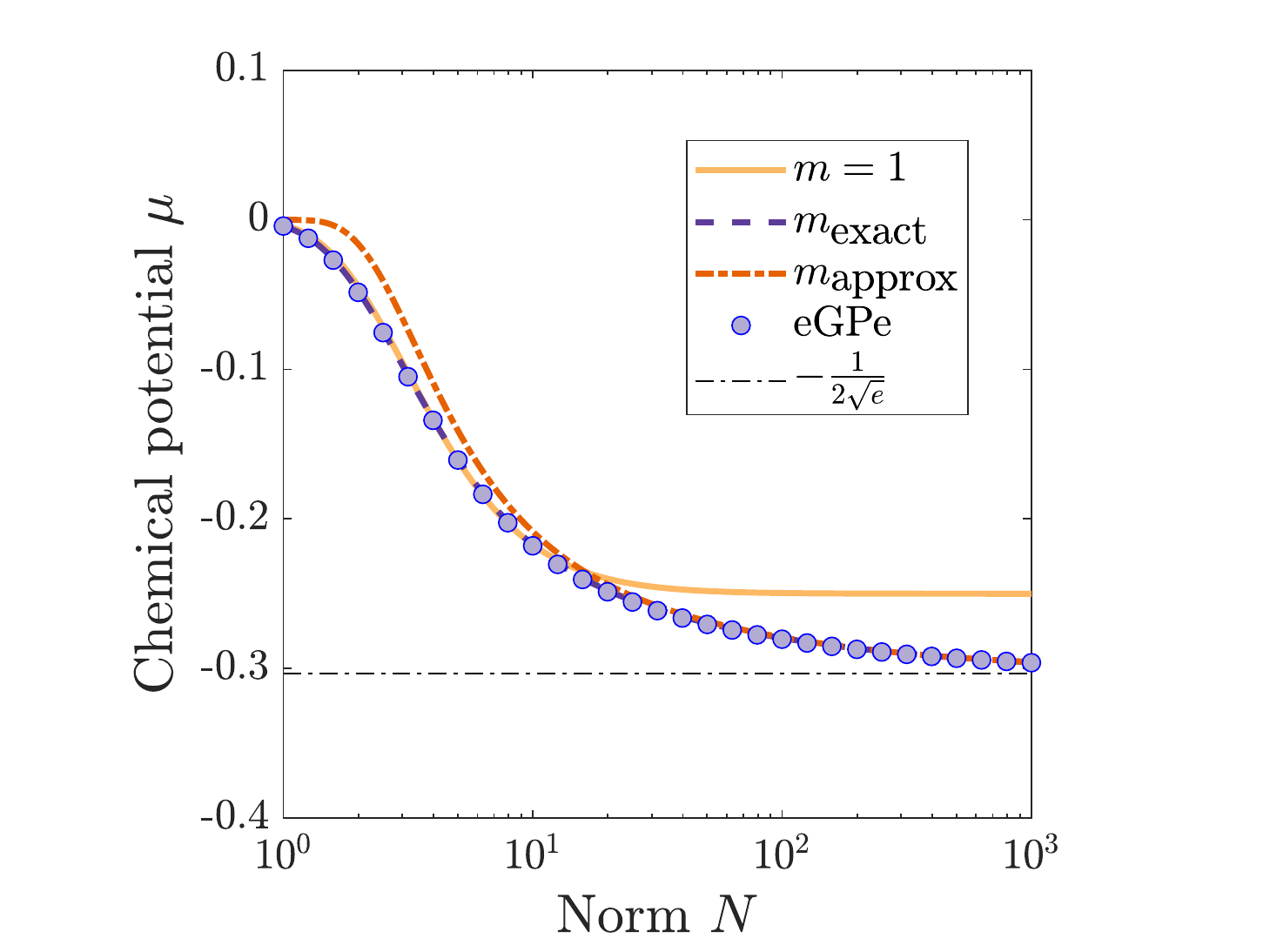}
    \caption{Chemical potential $\mu$ of the droplet based on the variational and numerical solutions, as specified by the legend.   The super-Gaussian ansatz correctly predicts the large $N$ limit, while the Gaussian ansatz clearly deviates from it. For small $N$ the approximate approach overestimates the chemical potential, while the Gaussian offers a good fit.}
    \label{fig:no_rot_omega-mu}
\end{figure}

We now compare the analytical results for the breathing mode frequency $\omega_0$ in Eq.\,(\ref{eq:var:freq_SG})  to the numerical results obtained by solving the eGPe, Eq.\,(\ref{eq:re:td-GPe}), for  $1\le N\le 1000$. We excite the breathing mode by introducing a small interaction perturbation, multiplying the interaction term in  Eq.\,(\ref{eq:re:td-GPe}) with a (dimensionless) factor $k=1.001$ which is decreased  back to unity linearly in a time interval $t_0 = 500$. The interaction perturbation $k$ is chosen such that the size modulations during the time propagation are small compared to $\sqrt{\langle r^2\rangle}$. The breathing frequency $\omega_0$ is measured via Fourier-analysis of $\sqrt{\langle r^2\rangle}$. We then calculate $\omega_0/|\mu|$ via the analytical chemical potential in Eq.\,(\ref{eq:var:chemical-pot_SG}) and Fig. \ref{fig:no_rot_omega-mu}, and also compare to the numerical solution. \\
\begin{figure}
    \centering
    \includegraphics[width=0.45\textwidth]{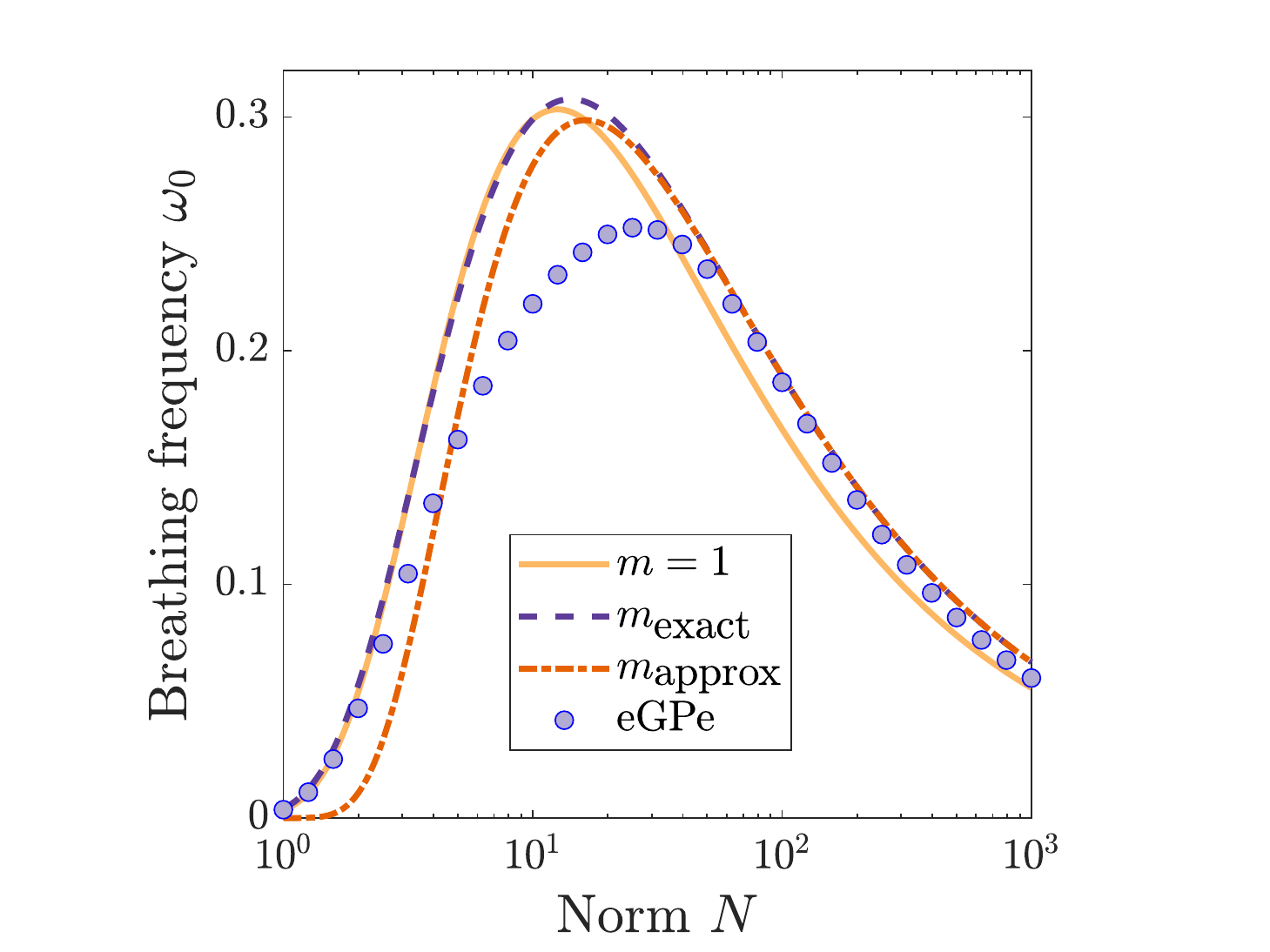}
    \caption{Breathing frequency $\omega_0$ of the non-rotating droplet. For large $N$ the numerical solutions of the eGPe, Eq.\,(\ref{eq:re:td-GPe}), coincide with the super-Gaussian ansatz calculation, while for smaller $N$ the assumption of a single breathing frequency in Eq.\,(\ref{eq:var:freq_eigenval}) no longer holds, due to the emergence of a beating pattern.}
    \label{fig:no_rot_omega-freq}
\end{figure}

Fig.~\ref{fig:no_rot_omega-freq} shows $\omega_0$ in comparison to analytical results with 
varying $m$, and in comparison to the numerical solution of the eGPe.
For large $N$ the approximate and exact solution coincide and give the correct behavior for $\omega_0$ as we can confirm from the comparison to the numerical solution of the eGPe. However, when approaching the low norm regime the numerical results start to deviate strongly from the analytical solutions. This is explained by an emerging beating pattern in the numerical results, such that there are several significant breathing mode frequencies. We assume however in Eq.\,(\ref{eq:var:freq_eigenval}) that the oscillating system only produces one single frequency. Similar to our observation for a two-dimensional droplet, the deviations between the Gaussian ansatz and the numerical results of the eGPe have also been found for a three-dimensional system~\cite{ref:col-ex_3D_huhui}. However, such deviations between analytical and numerical approaches are absent for the one-dimensional self-bound system~\cite{ref:col-ex-drop_astrakharchik}.
\begin{figure}
    \centering
    \includegraphics[width=0.45\textwidth]{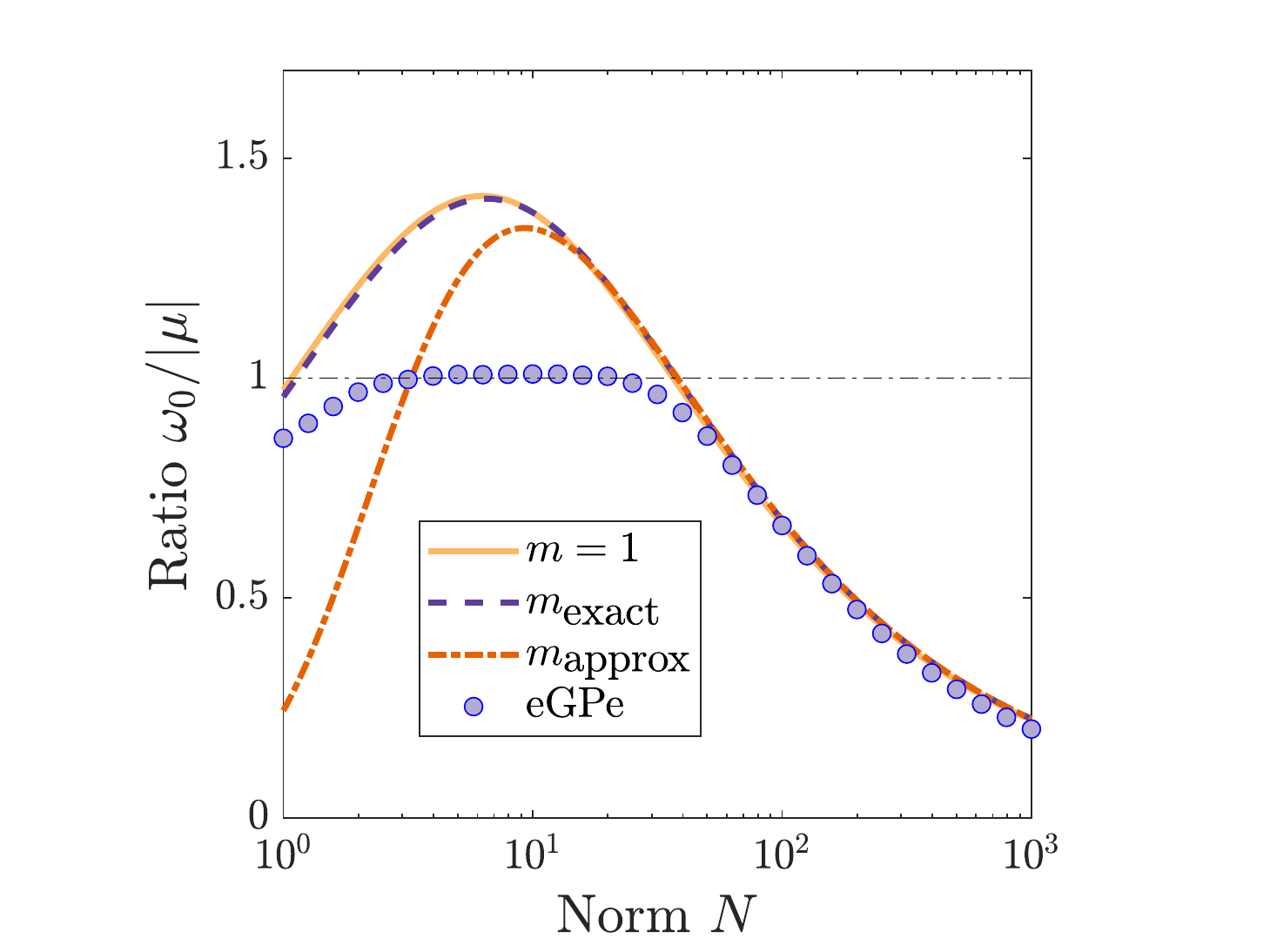}
    \caption{$\omega_0/|\mu|$ for droplets based on the different approaches, as specified in the legend.  All three analytical solutions coincide for the high norm regime, and are energetically favorable over the self-evaporation for $N\gtrsim36$.}
    \label{fig:no_rot_omega/mu}
\end{figure}
With the chemical potential in Eq.\,(\ref{eq:var:chemical-pot_SG}) we calculate the ratio $\omega_0/|\mu|$ shown in Fig. \ref{fig:no_rot_omega/mu}. Surprisingly, all three analytical solutions overlap for high $N$, while the eGPe solution starts to deviate as $\omega_0/|\mu|\rightarrow1$.
From the analytical results, we obtain that the breathing mode is observable from $N\gtrsim36$, while the numerical solution of the eGPe suggests $N\gtrsim21$. 


\section{Breathing mode in droplets with angular momentum\label{ch:numerical_results}}

\begin{figure}
    \centering
    \includegraphics[width=0.45\textwidth]{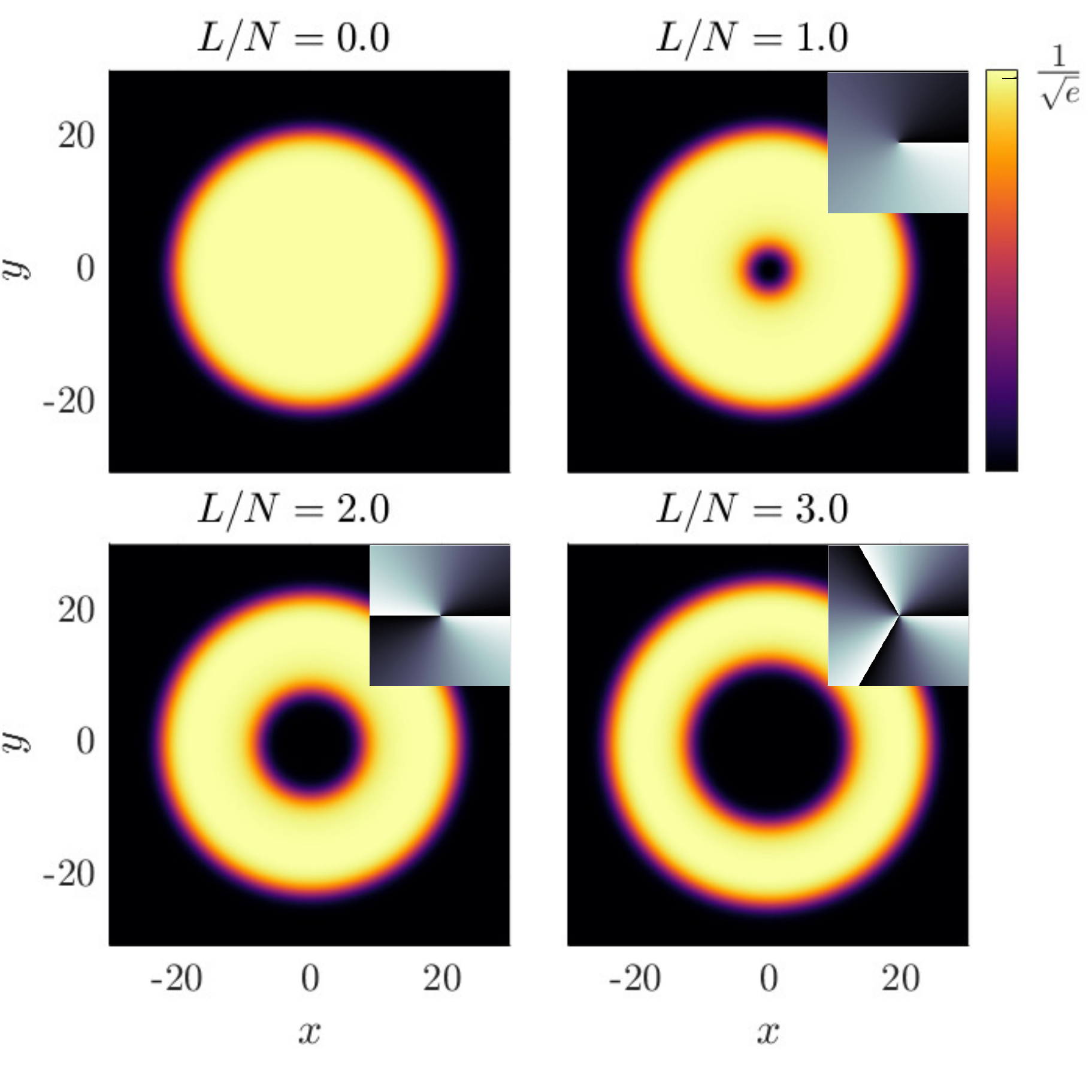}
    \caption{Density distributions for self-bound droplets at $N=850$ with angular momentum $L/N = 1.0,$ $2.0,$ $3.0$. The size of the droplets and minimum norm $N$ for stability follow the heuristically derived relation in Ref. \cite{ref:contact-drop_vortex_li}. The phase in cases with non-zero angular momentum is shown as small insets in each figure.}
    \label{fig:multi_charged}
\end{figure}

We continue our analysis by studying systems with angular momentum, obtained numerically by starting from initial states of the form
\begin{align}
\psi_0(\mathbf{r}) = Cr^S\exp{(-\alpha r^2+iS\theta)},
\end{align}
where $C$ is a normalization constant, $r$ and $\theta $ radial and angular coordinate, $\alpha$ some positive number, and $S$ is the vorticity~\cite{ref:contact-drop_vortex_li}.
Starting the time evolution from the states obtained this way, we then bring the interaction strength back to unity and obtain the breathing mode frequency $\omega_0$ via a Fourier-analysis of $\sqrt{\langle r^2\rangle}$ and the chemical potential $\mu$. Here, we choose systems with $L/N = 1.0,\,2.0,\,3.0$ as shown in Fig. \ref{fig:multi_charged} for $N=850$.
\begin{figure}
    \centering
    \includegraphics[width=0.45\textwidth]{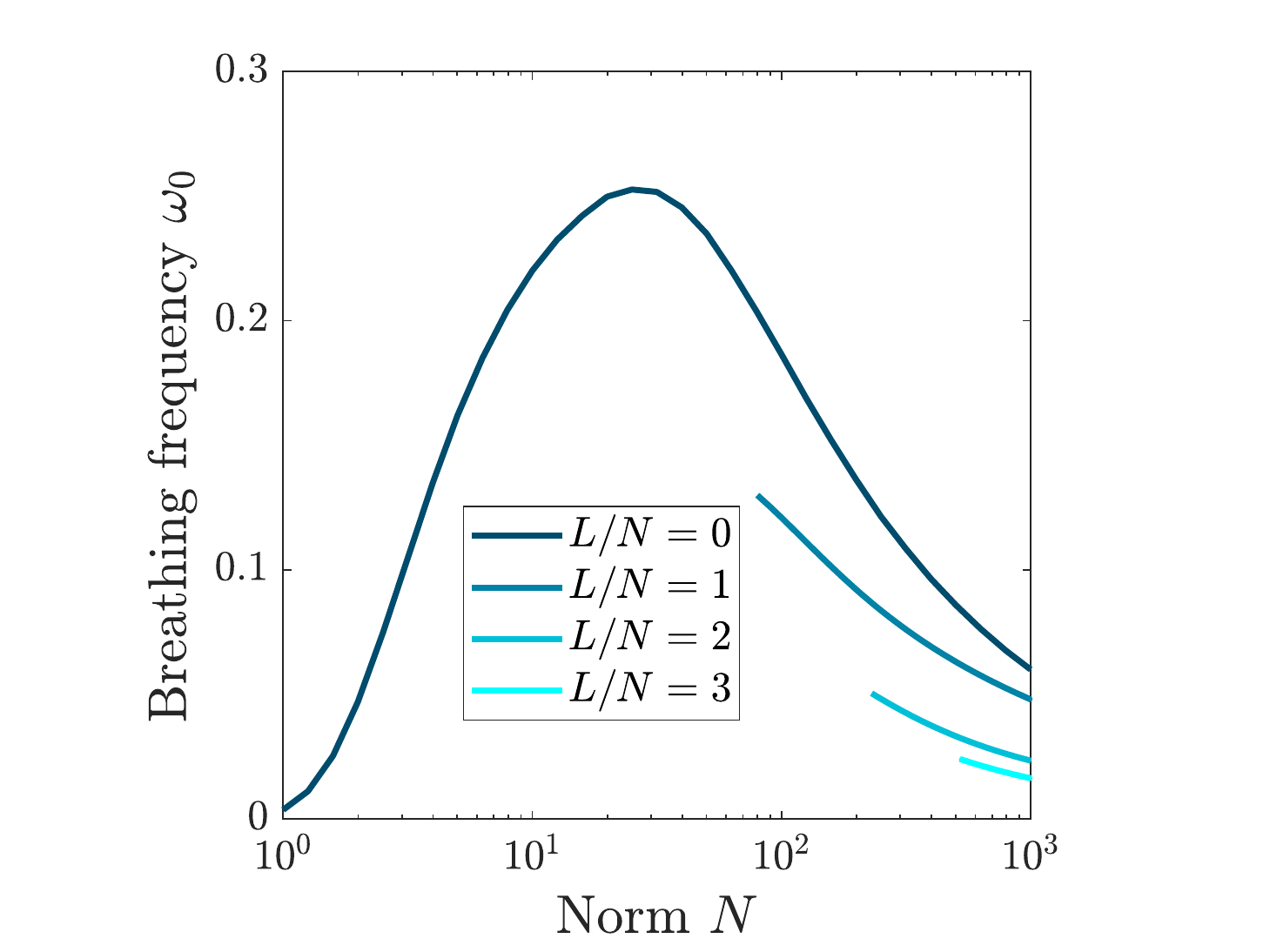}
    \caption{Frequency spectrum for droplets with angular momentum as specified in the legend. With increasing $L/N$ the breathing frequency $\omega_0$ decreases. The onset of each branch is given by the real-time stability of the system. For the multiply quantized systems, the stability requirement for the minimum required norm $N$ is equal to that found in \cite{ref:contact-drop_vortex_li}.}
    \label{fig:vortex_freq}
\end{figure}
\begin{figure}
    \centering
    \includegraphics[width=0.45\textwidth]{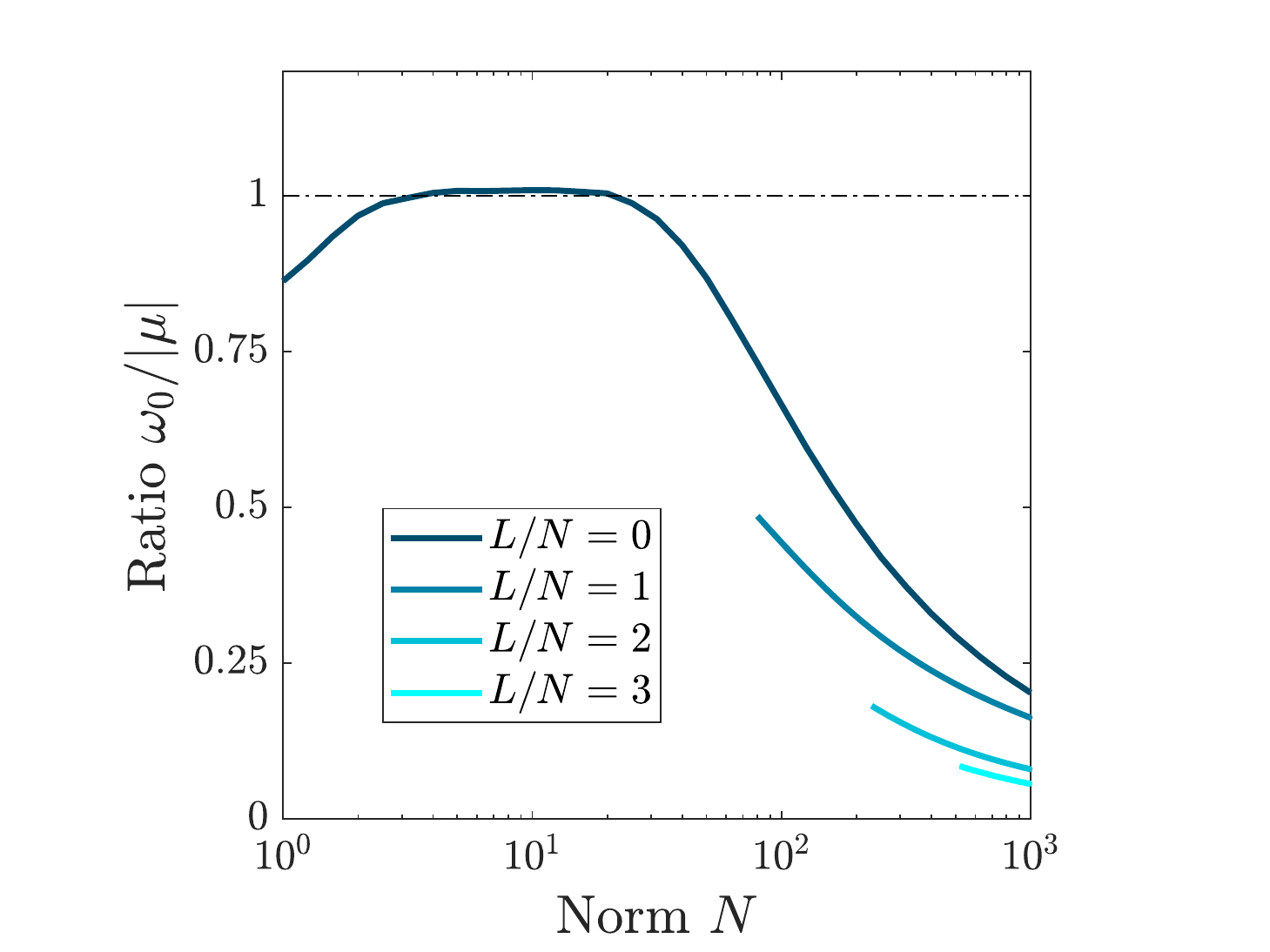}
    \caption{Availability of the breathing mode for droplets with angular momentum as given in the figure. The breathing mode exists if $\omega_0<|\mu| $ and is energetically favorable for a rotating system as long as the stability requirement is fulfilled and occurs for $L/N=0$ if $N\gtrsim21$.}
    \label{fig:vortex_freq_mu}
\end{figure}

As stated before, the breathing mode will only be observable if the condition
$\omega_0<\mu$ is fulfilled.
Due to the small oscillations of $\sqrt{\langle r^2\rangle}$ when 
measuring $\omega_0$ we can calculate $\mu$ numerically at any given 
point in the real-time evolution.
Figs.~\ref{fig:vortex_freq} and \ref{fig:vortex_freq_mu} show the breathing frequency $\omega_0$ and the ratio $
\omega_0/|\mu|$ for the systems as described above in the range $1\le N\le 1000$. Both quantities, $\omega_0$ and $\omega_0/|\mu|$ follow the same pattern, where the droplet without angular momentum corresponds to  the highest lying branch followed by the multiply-quantized vortices with lower $\omega_0$ as $L/N$ increases. 

With increasing $L/N$ the droplet requires a higher norm $N$ in order to support the size of the vortex core, as otherwise the modified kinetic energy due to angular momentum overcomes the attractive two-dimensional LHY term. We observe this behavior in Fig.~\ref{fig:vortex_freq} and \ref{fig:vortex_freq_mu} as the onset of the branches for the respective $L/N$ values. Further the onset follows the minimum $N$ requirement for a droplet to support a vortex with $L/N$ as in Ref.~\cite{ref:contact-drop_vortex_li}. As the numerical solution is the same as in Fig.~\ref{fig:no_rot_omega/mu}, the breathing mode continues to be observable from $N\gtrsim21$ while the analytical solution in Fig.~\ref{fig:no_rot_omega/mu} becomes observable for $N\gtrsim36$. This is in stark contrast to the one-dimensional case where the breathing mode is observable at any norm $N$ \cite{ref:contact-drop_col-ex_BdG_astrakharchik} and the quasi one-dimensional case \cite{ref:contact-drop-sol-cross_cappelaro} where it is observable for most configurations.


\section{Conclusions\label{ch:Conclusion}}

\noindent We have studied the breathing mode for symmetric two-dimensional self-bound binary Bose-gas droplets with components in the non-rotating case as well as for systems with a multiply-charged vortex imprinted at the center. For the breathing mode of the droplet with $L/N = 0$ we employed a variational super-Gaussian ansatz~\cite{ref:contact-drop_one-dim_superGauss} and compared it to the widely used Gaussian ansatz~\cite{ref:col-ex-drop_astrakharchik,ref:contact-drop-sol-cross_cappelaro,ref:col-ex_3D_huhui} as well as the numerical solution of the eGPe. We found that the super-Gaussian supersedes the Gaussian in predictability of stationary and dynamical properties of the self-bound system. However, for lower norms there is a great discrepancy between the numerical solution of the eGPe and the analytical solutions for the breathing mode, such that the ratio $\omega_0/|\mu|$ barely exceeds $1$ when the analytical solution suggests that $\omega_0/|\mu|>1$. Unlike two- and three-dimensional droplet~\cite{ref:col-ex_3D_huhui} systems, such discrepancy between the numerical and analytical solutions is not obtained for the one-dimensional system~\cite{ref:col-ex-drop_astrakharchik}. This suggests that additional considerations should be taken into account for the real-time evolution of the higher dimensional weakly interacting system within the numerical eGPe framework, and can be subject for future studies. 
We furthermore found that the breathing mode in self-bound droplets without angular momentum can be observed from $N\gtrsim36$, for which $\omega_0/|\mu|<1$. This is similar to the quasi one-dimensional and three-dimensional case~\cite{ref:contact-drop_3dim_Petrov,ref:contact-drop-sol-cross_cappelaro,ref:col-ex_3D_huhui}. However this behavior is different from that of a one-dimensional system, for which the breathing mode is observable at any $N$ \cite{ref:col-ex-drop_astrakharchik}.\\
We continued our investigation by introducing angular momentum in the shape of a multi-charged vortex imprinted at the droplet center. For this metastable system~\cite{ref:contact-drop_vortex_mikael} we find that with increasing $L/N$ the breathing frequency decreases, and droplets with angular momentum have $\omega_0/|\mu|<1$. Furthermore, the stability requirement in $N$ for droplets to support a vortex with $L/N$ is the same as in~\cite{ref:contact-drop_vortex_li}. 
The current work may be continued by further investigating the discrepancy between the numerically obtained breathing frequencies and the analytical solution. The super-Gaussian may be used as a trial order parameter in calculating the ideal trap-opening mechanism, while retaining the droplet state, commonly referred to as a shortcut to adiabaticity \cite{ref:STA_rashi,ref:STA_review}.
\bigskip

\begin{acknowledgments}
{\it Acknowledgements.} Discussions with J. Bengtsson and A. Idini are gratefully acknowledged. This work was financially supported by the Knut and Alice Wallenberg Foundation, the Swedish Research Council and NanoLund. 
\end{acknowledgments}

\bibliography{Sturmer_Dec2020.bib}

\end{document}